\documentstyle[12pt,amsbsy]{article}
\newcommand{\beq}{\begin{equation}}
\newcommand{\eeq}{\end{equation}}

\newcommand{\lsim}{\mbox{$<$\hspace{-0.8em}\raisebox{-0.4em}{$\sim$}}}

\begin{document}

\begin{titlepage}

\vspace{1cm}

\begin{center}
{\large How Is the Maximum Entropy of a Quantized Surface\\
\vspace{2mm} Related to Its Area?}
\end{center}

\begin{center}
I.B. Khriplovich\footnote{khriplovich@inp.nsk.su} and R.V.
Korkin\footnote{rvkorkin@mail.ru}
\end{center}
\begin{center}
Budker Institute of Nuclear Physics\\
630090 Novosibirsk, Russia,\\
and Novosibirsk University
\end{center}

\bigskip

\begin{abstract}
The maximum entropy of a quantized surface is demonstrated to be
proportional to the surface area in the classical limit. The
result is valid in loop quantum gravity, and in a somewhat more
general class of approaches to surface quantization. The maximum
entropy is calculated explicitly for some specific cases.
\end{abstract}

\vspace{8cm}

\end{titlepage}

We consider the relation between the maximum entropy of a
quantized surface and the area of this surface. To be definite, we
start the investigation with the approach to the surface
quantization based on loop quantum gravity (LQG)~[1-5]. Here, a
surface geometry is determined by a set of $\nu$ punctures on this
surface. In general, each puncture is supplied by two integer or
half-integer ``angular momenta'' $j^u$ and $j^d$:
\beq\label{j}
j^u,\, j^d= 0, 1/2, 1, 3/2, ...\;.
\eeq
$j^u$ and $j^d$ are related to edges directed up and down the
normal to the surface,  respectively, and add up into an angular
momentum $j^{ud}$:
\beq\label{ud}
{\bf j}^{ud}= {\bf j}^{u} + {\bf j}^{d}; \quad |j^{u}-j^{d}|\leq
j^{ud} \leq j^{u}+j^{d}.
\eeq
The area of a surface is
\beq\label{Aj}
A =\alpha\, l_p^2 \sum_{i=1}^{\nu}\sqrt{2 j^u_i(j^u_i+1)+
2j^d_i(j^d_i+1)- j^{ud}_i(j^{ud}_i+1)}\;.
\eeq

A comment on the last formula is appropriate here. It is quite
natural that the Planck length squared
\beq
l_p^2 = {G\hbar \over c^3}
\eeq
serves as the unit area. The generalized quantum number
\beq\label{N}
N=\sum_{i=1}^{\nu}\sqrt{2 j^u_i(j^u_i+1)+ 2j^d_i(j^d_i+1)-
j^{ud}_i(j^{ud}_i+1)}\;,
\eeq
is on the order of $j^u_i$ and $j^d_i$. Then, for $A$ to be finite
in the classical limit of a large sum of quantum numbers, the
power of $N$ should be equal to that of $\hbar$ in $l_p^2$. This
argument, formulated in \cite{khr}, is also quite natural. It can
be checked easily by inspecting any expectation value nonvanishing
in the classical limit in ordinary quantum mechanics. This is why
just $N$, i. e. the sum of square roots, but not for instance the
sum of $j(j+1)$ or of $\sqrt[4]{j(j+1)}$, should enter the
expression for area.

As to the overall numerical factor $\alpha$ in (\ref{Aj}), it
cannot be determined without an additional physical input. This
ambiguity originates from a free (so-called Immirzi) parameter
\cite{imm,rot} which corresponds to a family of inequivalent
quantum theories, all of them being viable without such an input.
One may hope that the value of this factor in (\ref{Aj}) can be
determined by studying the entropy of a black hole. This idea
(mentioned previously in~\cite{boj}) was investigated by one of
us~\cite{khri} for rather simplified models under the assumption
that the entropy of an eternal black hole in equilibrium is
maximum. This assumption goes back to~\cite{vaz}, where it was
used in a model of the quantum black hole as originating from dust
collapse. In the present article we confine to the calculation of
the maximum entropy of a surface with the area given by relation
(\ref{Aj}) (or some its generalization).

The entropy $S$ of a surface is defined as the logarithm of the
number of states of this surface with fixed area $A$, i.e. fixed
sum (\ref{N}). Let $\nu_i$ be the number of punctures with a given
set of momenta $j^u_i$, $j^d_i$, $j^{ud}_i$. The total number of
punctures is
\[
\nu = \sum_i \nu_i.
\]
To each puncture $i$ one ascribes a statistical weight $g_i$.
Since ${\bf j}^{ud}_i= {\bf j}^{u}_i + {\bf j}^{d}_i$, this
statistical weight equals, in the absence of new constraints, to
the number of possible projections of ${\bf j}^{ud}_i$, i.e. $g_i
= 2 j^{ud}_i + 1$. Then the entropy is
\beq\label{en1}
S=\ln\left[\prod_i\,\frac{(g_i)^{\nu_i}}{\nu_i\,!} \nu\,! \right]
=\sum_i \nu_i \ln g_i - \sum_i \ln \nu_i\,! + \ln \nu\,!\,.
\eeq
The structure of expressions (\ref{Aj}) and (\ref{en1}) is so
different that naturally in a general case the entropy cannot be
proportional to the area (see a more detailed discussion
in~\cite{khri}). However, as will be demonstrated now, this is
the case for the maximum entropy in the classical limit.

By combinatorial reasons, it is natural to expect that the
absolute maximum of entropy is reached when all values of quantum
numbers $j_i^{u,d,ud}$ are present. This guess is confirmed as
well by concrete calculations for some model cases (see
\cite{khri}). We assume also that in the classical limit the
typical values of puncture numbers $\nu_i$ is large. Then, with
the Stirling formula for factorials, expression (\ref{en1})
transforms to
\beq\label{en2}
S= \sum_i \left[\nu_i \,\ln (2j+1) - \left(\nu_i
+\frac{1}{2}\right)\,\ln \nu_i \right]+ \left(\sum_i \nu_i +
\frac{1}{2}\right)\times \ln \left(\sum_{i^{\prime}}
\nu_{i^{\prime}}\right).
\eeq
We have omitted here terms with $\ln\sqrt{2\pi}$, each of them
being on the order of unity. The validity of this approximation,
as well as of the Stirling formula by itself for this problem,
will be discussed later.

We are looking for the extremum of expression (\ref{en2}) under
the condition
\beq\label{con}
N=\sum_i \nu_i\,r_i\, = {\rm const}, \quad {\rm where} \quad
r_i=\sqrt{2 j^u_i(j^u_i+1)+ 2j^d_i(j^d_i+1)-
j^{ud}_i(j^{ud}_i+1)}\;.
\eeq
The problem reduces to the solution of the system of equations
\beq\label{sys}
\ln g_i - \ln \nu_i +  \ln \left(\sum_{i^{\,\prime}}
\nu_{i^{\,\prime}}\right) = \mu r_i\,,
\eeq
or
\beq\label{nu}
\nu_i = g_i\, e^{- \mu r_i}\,\sum_{i^{\,\prime}}
\nu_{i^{\,\prime}}\,.
\eeq
Here $\mu$ is the Lagrange multiplier for the constraining
relation (\ref{con}). Summing expressions (\ref{nu}) over $i$, we
arrive at the equation on $\mu$:
\beq\label{equ}
\sum_i g_i\, e^{- \mu r_i}= 1.
\eeq
On the other hand, when multiplying equation (\ref{sys}) by
$\nu_i$ and summing over $i$, we arrive with the constraint
(\ref{con}) at the following result for the maximum entropy for a
given value of $N$:
\beq\label{enf}
S_{\rm max}= \mu \,N\,.
\eeq
Here the terms
\[
-\,\frac{1}{2}\sum_i \ln \nu_i \quad \mbox{\rm and} \quad
\frac{1}{2}\ln \nu
\]
in the expression (\ref{en2}) have been neglected. Below we will
come back to the accuracy of this approximation.

Thus, it is the maximum entropy of a surface which is
proportional in the classical limit to its area. This
proportionality certainly takes place for a classical black hole.
And this is a very strong argument in favour of the assumption
that the black hole entropy is maximum.

It should be emphasized that relation (\ref{enf}) is true not only
in LQG, but applies to a more general class of approaches to the
quantization of surfaces. What is really necessary here, is as
follows. The surface should consist of patches of different sorts,
so that there are $\nu_i$ patches of each sort $i$, with a
generalized effective quantum number $r_i$, and a statistical
weight $g_i$. Then in the classical limit, the maximum entropy of
a surface is proportional to its area.

As a simple example, let us consider the situation when the
general formula (\ref{Aj}) reduces to
\beq\label{A1}
A =\alpha\, l_p^2 \sum_{i=1}^{\nu} \sqrt{j_i(j_i+1)}\;=\alpha\,
l_p^2 \sum_{j=1/2}^{\infty}\sqrt{j(j+1)}\;\nu_j\,.
\eeq
Previously \cite{khri}, the entropy was found in this case under
the assumption of the fixed sum of quantum numbers
\[
n=\sum_{j=1/2}^{\infty} j\,\nu_j\,.
\]
Now we will solve this problem for fixed
\beq\label{N1}
N=\sum_{j=1/2}^{\infty}\sqrt{j(j+1)}\;\nu_j\,.
\eeq
Here the statistical weight of a puncture with the quantum number
$j$ is $g_j= 2j + 1$, and equation (\ref{equ}) can be rewritten as
\beq\label{eq1}
\sum_{p=1}^{\infty}(p+1)\,z^{\sqrt{p(p+1)}}=1, \quad p=2j, \quad
z=e^{-\mu/2}.
\eeq
Its solution is
\beq\label{z1}
z=0.423, \quad \mbox{or} \quad \mu=-2\ln z = 1.722\,.
\eeq
Thus, here the maximum entropy is
\beq\label{s1}
S_{\rm max}=1.722N=2.515\nu.
\eeq
The mean value of the angular momentum is
\beq\label{j1}
<j>=\frac{1}{\nu}\sum_{j=1/2}^{\infty} j \, \nu_j = 1.059.
\eeq

In a sense, the simplest choice for the quantum numbers $j_i$ in
this model is to put all of them equal to $1/2$. Then $\nu_j =\nu
\delta_{j,1/2}$, and
\beq\label{s}
S=\ln2\,\nu\,, \quad \mbox{\rm or} \quad S=\frac{2 \ln
2}{\sqrt{3}}\,N\,.
\eeq
In fact, this is the value of the black hole entropy derived
previously in \cite{asht} within a Chern-Simons field theory; the
typical value of $j_i$ obtained therein is also $1/2$.

One more example is as follows. Here at each puncture we have
$j^u_i= j^d_i = j_i, \quad j^{ud}_i =0$. In this case
\beq\label{N2}
N=2\sum_{j=1/2}^{\infty}\sqrt{j(j+1)}\;\nu_j\,,
\eeq
and $g_j=1$. Then the equation for $z=\exp(-\mu/2)$ is
\beq\label{eq2}
\sum_{p=1}^{\infty} z^{\sqrt{p(p+1)}}=1\,,
\eeq
with the solution
\beq\label{z2}
z=0.602, \quad \mbox{or} \quad \mu=-2\ln z = 0.508\,.
\eeq
Thus obtained maximum entropy and the mean angular momentum are
\beq\label{s2}
S_{\rm max}=0.508 N=1.655 \nu,
\eeq
\beq\label{j2}
<j>= 1.224.
\eeq

Let us consider at last the general case, with $N$ given by
formula (\ref{N}), $g_i= 2j^{ud}_i +1$, and with all values of
$j^u_i$, $j^d_i$, $j^{ud}_i$ allowed. In this case the solution
of equation (\ref{equ}) is
\beq\label{z3}
z=0.202\,,
\eeq
and the maximum entropy equals
\beq\label{s3}
S_{\rm max}=3.120 N=4.836 \nu.
\eeq
The mean values of quantum numbers are
\beq\label{j3}
<j^u>=<j^d>= 1.072, \quad <j^{ud}>=2.129.
\eeq

It should be emphasized that in this way one always arrives at the
quantization rule for the black hole entropy (and area)
effectively with integer quantum numbers $\nu$, as proposed in the
pioneering article~\cite{bek} (see also \cite{muk,kog}).

Let us discuss now the accuracy of our result for the maximum
entropy. To make our arguments more concrete and clear, we will
consider the second of the above models, with $N$ given by formula
(\ref{eq2}), and results described by (\ref{z2})-(\ref{j2}).
However, one can readily check that the estimates obtained here
are valid qualitatively both for the first model described by
formulae (\ref{N1})-(\ref{j1}), and for the third, most general
case. With $<j> \, \simeq 1$, the number of punctures $\nu$ is on
the same order of magnitude as $N$. Thus, in the classical limit,
$\nu \sim N \gg 1$. Now, according to relation (\ref{nu}), with
$z \, \lsim \,1$, the numbers $\nu_j$ satisfy the condition
$\nu_j > 1$ as long as the quantum numbers $j$ are bounded as
follows:
\beq\label{lsim}
j \,\lsim \,J \sim \ln N.
\eeq
Clearly, the typical values of those $\nu_j$ which give essential
contributions to $N$ are large, and the Stirling approximation for
$S$ is fully legitimate.

On the other hand, the number of terms in the sums over $j$ in the
expressions of the type (\ref{en2}) is effectively bounded by
inequality (\ref{lsim}). Thus, the contribution of the terms with
$\ln\sqrt{2\pi}$, omitted in (\ref{en2}), as well as of the term
\[
\frac{1}{2}\ln \nu
\]
retained in (\ref{en2}), but neglected in the final expression
(\ref{enf}), is on the order of $\ln N$ only. The leading
correction to our result (\ref{enf}) originates from the term
\[
-\,\frac{1}{2}\sum_i \ln \nu_i\,,
\]
and constitutes $\sim \ln^2 N$ in order of magnitude. By the way,
this leading correction is different from that for the model
considered in \cite{kama,car}, where this correction is on the
order of $\ln N$.

In conclusion, let us mention the attempts made in \cite{rove,kra}
to calculate the surface entropy in LQG. In those papers the
distribution of the angular momenta $j$ over the edges is not
discussed at all. We cannot understand how one could find the
surface entropy without such information.

\bigskip
\bigskip
\begin{center}***\end{center}
We are grateful to G.G. Kirilin for the interest to the work and
useful discussions. The investigation was supported in part by
the Russian Foundation for Basic Research through Grant No.
01-02-16898, through Grant No. 00-15-96811 for Leading Scientific
Schools, by the Ministry of Education Grant No. E00-3.3-148, and
by the Federal Program Integration-2001.

\end{document}